\documentclass[epj]{svjour}
\usepackage{graphics}
\begin{document}

\title{Memory interference effects in spin glasses}
\author{
K. Jonason\inst{1,2}, P. Nordblad\inst{1}, E. Vincent\inst{2},
J. Hammann\inst{2}, 
J.P. Bouchaud\inst{2}
}                     
\offprints{Per.Nordblad@Angstrom.uu.se or vincent@spec.saclay.cea.fr}          
\institute{Dept. of Materials Science, Uppsala University, Box 534,
SE-751 21 Uppsala, Sweden 
\and 
Service de Physique de l'Etat Condens\'e, CEA Saclay,
91191 Gif sur Yvette Cedex, France }
\date{Received: April 28, 1999/ Revised version: June 17, 1999}
%
\abstract{
When a spin glass is cooled down, a memory of the cooling process is
imprinted in the spin structure.  This memory can be disclosed in a
continuous heating measurement of the ac-susceptibility.  E.g., if a
continuous cooling process is intermittently halted during a certain
aging time at one or two intermediate temperatures, the trace of the
previous stop(s) is recovered when the sample is continuously
re-heated \cite{uppsalaSaclay}.  However, heating the sample above the
aging temperature, but keeping it below $T_{g}$, erases the memory of
the thermal history at lower temperatures.  We also show that a memory
imprinted at a higher temperature can be erased by waiting a long
enough time at a lower temperature.  Predictions from two
complementary spin glass descriptions, a hierarchical phase space model and
a real space droplet picture are contested with these memory
phenomena and interference effects.
\vskip 0.5cm
\noindent
[1] K. Jonason, E. Vincent, J.
Hammann, J. P. Bouchaud and P. Nordblad, {\it Phys.  Rev.  Lett.}  {\bf 31},
3243 (1998).
\vskip 0.2cm
} 
\maketitle
%
\section{Introduction}
\label{intro}

 The non-equilibrium character of the dynamics of 3d spin glasses
below the zero field phase transition temperature has been extensively
studied by both experimentalists and theorists \cite{sgrf}.  Two
different main tracks have been used to describe the aging and
non-equilibrium dynamics that is characteristic of spin glasses.  On
the one hand phase space pictures \cite{hierarki}, which originate
from mean field theory \cite{mpv} and prescribe a hierarchical
arrangement of metastable states, and on the other hand real space
droplet scaling models which have been developed from renormalisation
group arguments \cite{FH,Henk}. Independently, a theoretical
description (which we shall not discuss here) of aging effects
\cite{CuKu1} and temperature variation effects \cite{CuKu2} has been
given by the direct solution of the dynamical equations of mean-field
like models.

When a spin glass is quenched from a high temperature (above $T_{g}$)
to a temperature $T_{1}$ below $T_{g}$, a wait time dependence of the
dynamic magnetic response is observed.  This aging behaviour
\cite{aging,Sitges} corresponds to a slow evolution of the spin
configuration towards equilibrium.  The `magnetic aging' observed
in spin glasses resembles the `physical aging' observed in the
mechanical properties of glassy polymers \cite{polymer,cavaille}, or
the `dielectric' aging found in supercooled liquids \cite{nagel}
and dielectric crystals \cite{KTNKLT}.  However, a more detailed
comparison of aging in these various systems would show interesting
differences \cite{nagel}, particularly with respect to the effect of
the cooling rate \cite{uppsalaSaclay}.

Remarkable influences of slight temperature variations on the aging
process in spin glasses have been evidenced in a wide set of earlier
experiments \cite{hierarki,Sitges,UppsDT,DjurCuMn}.  These influences
were further elucidated through the memory phenomena recently reported
in \cite{uppsalaSaclay}.  The results have been interpreted both from
`phase space' and `real space' points of view.

In phase space models, aging is pictured as a random walk among the
metastable states.  At a given temperature $T_{1}$, the system samples
the valleys of a fixed free-energy landscape.  On the basis of the
experimental observations, it has been proposed \cite{hierarki} that
the landscape at $T_1$ corresponds to a specific level of a
hierarchical tree.  When lowering the temperature to $T_{2}<T_1$, the
observed restart of aging is explained as a subdivision of the free
energy valleys into new ones at a lower level of the tree.  The system
now has to search for equilibrium in a new, unexplored landscape, and
therefore acts at $T_2$ as if it had been quenched from a high
temperature.  On the other hand, the experiments show that when
heating back from $T_{2}$ to $T_{1}$ the memory of the previous aging
at $T_{1}$ is recovered.  In the hierarchical picture, this is
produced by the $T_2$-valleys merging back to re-build the
$T_1$-landscape.

In real space droplet pictures \cite{FH,Henk}, the aging behaviour at
constant temperature is associated with a growth of spin glass ordered
regions of two types (related by time-reversal symmetry).  This is
combined with a chaotic behaviour as a function of temperature,
\cite{braymoore}, i.e.  the equilibrium spin configuration at one
temperature is different from the equilibrium configuration at another
temperature.  However, there is also an overlap between the
equilibrium spin structures at two different temperatures, $T$ and
$T\pm\Delta T$, on length scales shorter than the overlap length,
$L_{\Delta T}$.  In this picture, chaos implies that if the spin glass
has been allowed to age a time, $t_{w}$, at a certain temperature, the
aging process is re-initialized after a large enough temperature
change.  Intuitively, a growth of compact domains may not allow a
memory of a high temperature spin configuration to remain imprinted in
the system while the system ages at lower temperatures.  However, as
suggested in \cite{uppsalaSaclay} and developed in \cite{jonssonetal},
a phenomenology based upon fractal domains and droplet excitations can
be able to incorporate the observed memory behaviour in a
real space droplet picture.  The possibility of a fractal
(non-compact) geometry of the domains has been evoked in the past in
various theoretical contexts\cite{fractaldiv}, and also in close
connection with the aging phenomena \cite{fractclust,jpbdean,mfnoneq}.

In this paper, we report new results on the memory phenomenon observed
in low frequency ac-susceptibility of spin glasses.  We first recall
and demonstrate an undisturbed memory phenomenon, and then show that
such a memory can be erased not only by heating the sample to a
temperature above the temperature where the memory is imprinted, but
also by waiting a long enough time below this temperature.

\section{Experimental}
\label{expe}

The experiments were performed on the insulating spin glass
$CdCr_{1.7}In_{0.3}S_{4}$ \cite{mtrl}
($T_{g}=16.7 K$), in a Cryogenic Ltd S600 SQUID magnetometer at Saclay.  The ac
field used in the experiments had a peak magnitude of 0.3 Oe and
frequency $\omega/2\pi$=0.04 Hz.  This low frequency makes the
relaxation of the susceptibility at a constant temperature in the spin
glass phase clearly visible (at the laboratory time scale of $10^1$ to
$10^5$ s).

\begin{figure}
\resizebox{0.47\textwidth}{!}{%
\includegraphics{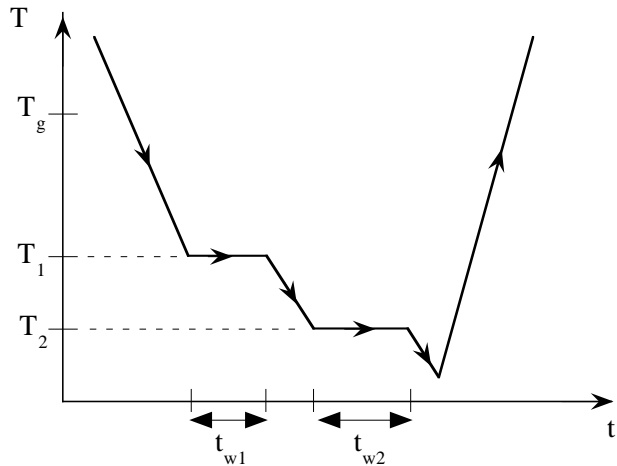} }
\caption{ The measurement procedure in the `double memory experiments'
of Fig.2, 4 and 5.}
\label{un}
\end{figure}

The basic experimental procedure is illustrated in Fig. 1 and is as follows:

(i) Cooling: The experiments are always started at 20 K, a temperature
well above the spin glass temperature $T_{g}$=16.7 K.  The
ac-susceptibility is first recorded as a function of decreasing
temperature.  The sample is continuously cooled, but is additionally
kept at constant temperature at two intermittent temperatures $T_{1}$
and $T_{2}$ for wait times $t_{w1}$ and $t_{w2}$, respectively 
($T_{1}<T_{2}<T_{g}$).

(ii) Heating: When the lowest temperature has been reached, the system
is immediately continuously re-heated and the ac-susceptibility is
recorded as a function of increasing temperature.

Except at $T_{1}$ and $T_{2}$ when decreasing the temperature, the
cooling and heating rates are constant ($\sim 0.1$ K/min.).  At
constant temperature, both components of the ac-susceptibility relax
downward by about the same absolute amount.  However, the relative
decay of the out-of-phase is much larger than the relative decay of
the in-phase component, and in the following we mainly focus on results
from the out-of-phase component of the susceptibility.

\section{Results}
\label{resu}

\subsection{Double memory}

The results of a double memory experiment are presented in Fig.  2.
The initial data is recorded on continuously cooling the sample
including a first halt at the temperature $T_{1}$=12 K (0.72$T_{g}$)
for $t_{w1}$=7 $hrs$ and a second halt at $T_{2}$=9 K (0.54$T_{g}$)
for $t_{w2}$=40 $hrs$.  The cooling is then continued to $T$=5 K,
from where a new set of data is taken on increasing the temperature at
a constant heating rate whithout halts.  A reference curve, measured
on continuous heating after cooling the sample without intermittent
halts, is included in the figure.

\begin{figure}
\vskip 2.1cm
\resizebox{0.48\textwidth}{!}{%
\includegraphics{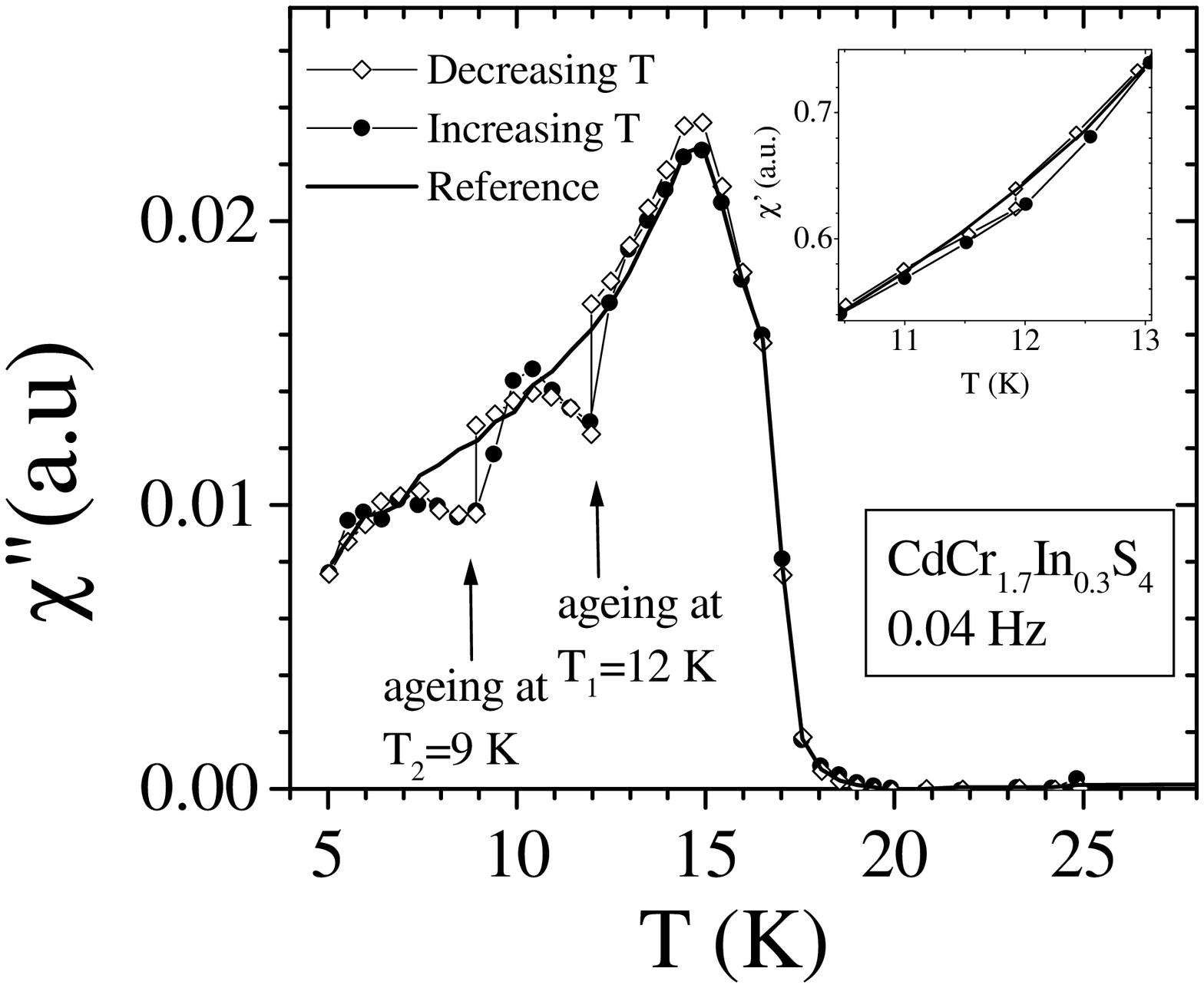} }
\caption{ Out-of-phase susceptibility vs. temperature. While cooling
down (open diamonds), two intermittent halts are made; at $T_{1}$=12
$K$ during $t_{w1}$=7 $hrs$, and at $T_{2}$=9 K during $t_{w2}$=40
$hrs$. The system is then reheated at a constant heating rate (full
circles). The reference curve (solid line) is measured on heating the
sample after cooling it without intermittent halts. The inset shows
the in-phase susceptibility in the same measurement procedure, at
temperatures around $T_{1}$=12 K.}

\label{deux}
\end{figure}

A first important feature can be noted on the curve recorded on
cooling with intermittent halts.  After aging 7 $hrs$ at 12 K, $\chi''$
has relaxed downward due to aging.  But when cooling resumes, the
curve rises and merges with the reference curve, as if the aging at 12 K
was of no influence on the state of the system at lower temperatures.
This chaos-like effect (in reference to the notion of chaos in
temperature introduced in \cite{braymoore}) points out an important
difference from a simpler description of glassy systems, in which
there are equivalent equilibrium states at all low temperatures and
aging at any temperature implies that this equilibrium state is
further approached.  Here, as pictured in more detail in other
experiments \cite{uppsalaSaclay}, only the last temperature interval
of the cooling procedure does contribute to the approach of the
equilibrium state at the final temperature.  

Note that this notion of
 `last temperature interval' depends on the observation time scale
of the measurement ($\chi''$ is only sensitive to dynamical processes
with a characteristic response time of order $1/\omega$ which in these
experiments corresponds to $\approx 4$ s).  In magnetisation relaxation
experiments, the observation time corresponds to the time elapsed after
the field change, and effects of aging at a higher temperature can be
seen in the long-time part ($10^3-10^4$ s) of the relaxation curves
\cite{marcos,DjurCuMn} in a correspondingly enlarged `last temperature
interval' compared to that of our current ac-susceptibility
experiments.

The curve recorded on re-heating in Fig.  2 clearly displays the
memory effect: the dips at $T_{1}$ and $T_{2}$ are recovered.  The
long wait time (40 $hrs$) at $T_{2}$= 9 K has no apparent influence on
the memory dip associated with $T_{1}$=12 K.  This experiment
displays a double memory where no interference effects are present,
i.e.  the two dips at $T_{1}$ and $T_{2}$ are, within our experimental
accuracy, fully recovered when reheating the sample.  A similar result
has been obtained on a metallic Cu:Mn spin-glass sample
\cite{uppsalaSaclay,DjurCuMn}, confirming the universality of aging
dynamics in very different spin-glass realizations. 

This experimental procedure has been recently reproduced in extensive
simulations of the 3d Edwards-Anderson model. Although weaker and
more spread out in temperature, similar effects of a restart of
aging upon cooling and of a memory effect upon heating have been
found \cite{Takayama}.

The memory phenomenon is also observable in the in-phase
susceptibility, $\chi$'.  In the inset of Fig.  2, $\chi$' is plotted
in the region around $T_{1}$=12 $K$.  A relaxation due to aging is
visible, and it is also clear that when cooling resumes after aging,
the $\chi$' curve rather rapidly merges with the reference curve
(chaos-like effect).  Upon re-heating, the memory effect can be
distinguished, the memory curve clearly departs from the reference in
the 11-13K range. The relative weakness of the deviation compared to
the large $\chi$' value can be understood in the following way. 
Aging in spin glasses mainly affects processes with relaxation times
of the order of the age of the system; processes with shorter
relaxation times are already equilibrated, and processes with longer
relaxation times are not active. $\chi$' measures the integrated
response of all short time processes up to the observation time
$1/\omega$, whereas $\chi$'' only probes processes with relaxation
times of order $1/\omega$. The relative influence of aging is thus
smaller in $\chi$' than in $\chi$''.

\subsection{Memory erasing by heating}

\begin{figure}
\vskip 2.0cm
\resizebox{0.49\textwidth}{!}{%
\includegraphics{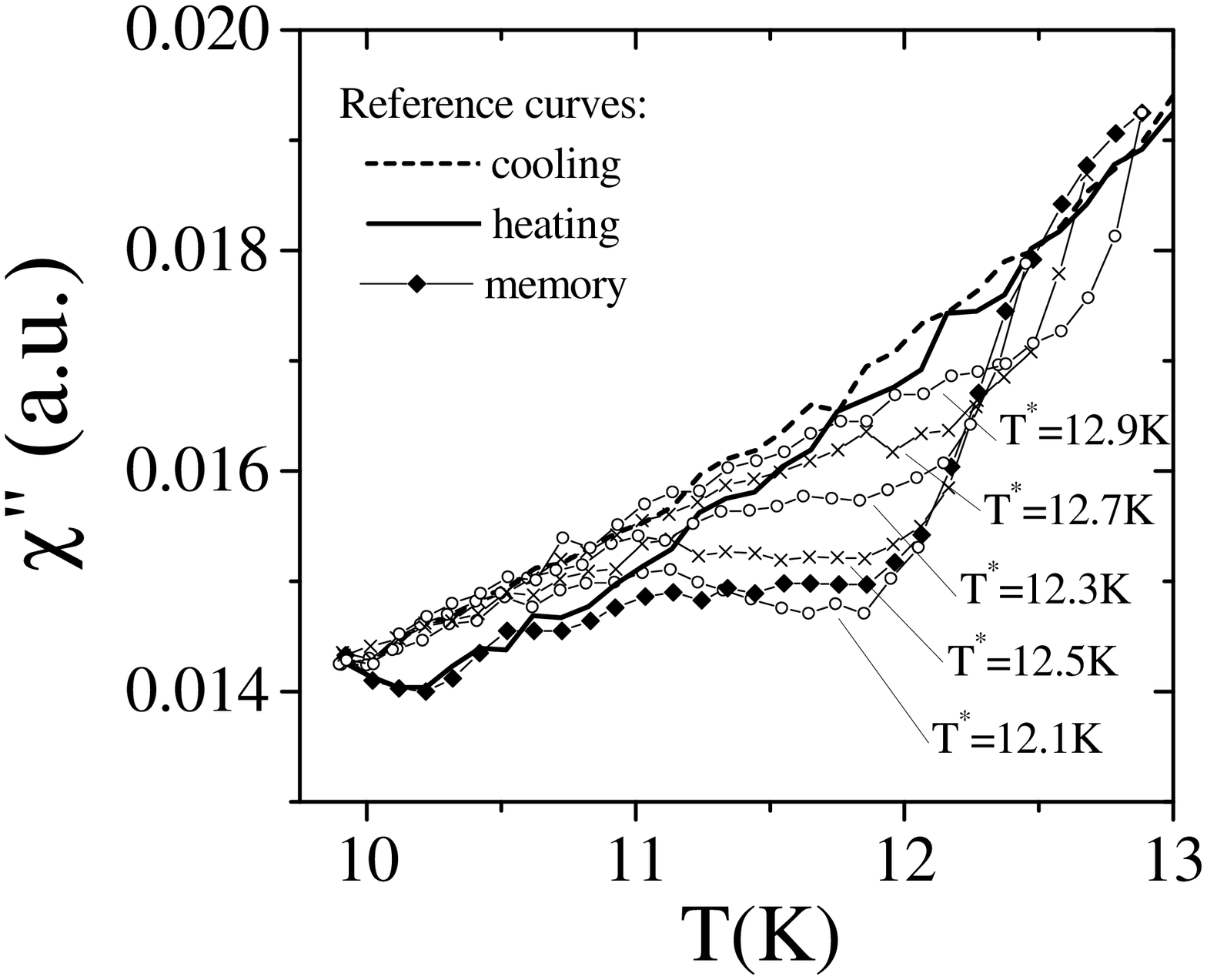} }
\caption{ Memory erasing by overheating.  The memory effect at 12K,
recorded during heating with the same procedure as in Fig.2, is shown
in full diamonds. Now, when reaching $T^*$ ($T^*$= 12.1, 12.3, 12.5,
12.7 and 12.9 K), re-heating stops and the sample is cooled back.  The
$\chi''$ signal (alternatively open circles and crosses for the
various $T^*$ values) during cooling shows the progressive erasing of
the 12K memory for increasing $T^*$.  Reference curves, measured
during a continuous cooling (dashed line) and heating (solid line),
are also shown.  }
\label{trois}
\end{figure}

The memory of aging at $T_1$ remains imprinted in the system during
additional aging stages at sufficiently lower temperatures, and is
recovered when heating back to $T_1$. We have  investigated what remains of this $T_1$-memory
after heating up to $T^*>T_1$ (keeping of course $T^*<T_g$).  The
experiments were performed using only one intermittent stop at
$T_1=12$ K for $t_{w1}$= 3 $hrs$, and continuing the cooling to about 10
K. Then
$\chi''$  was recorded upon heating the sample to $T^*$ and 
immediately re-cooling it to 10 K. The results are displayed in Fig.  3.
For higher and higher $T^*$, the memory dip at $T_1$ becomes weaker
and weaker, finally fading out at $T^*\sim$ 13 K. The additional
shallow dips observed in a limited temperature region just below $T^*$
and just above 10 K, the two temperatures where the temperature change
is reversed, are due to the finite heating/cooling rate and to the overlap within this temperature range between the state created
on heating (cooling) and the desireable state on re-cooling
(re-heating) the sample \cite{jonssonetal}.

\subsection{Memory interference}

The memory is also affected by aging at a lower temperature, provided
this temperature is close enough to $T_{1}$ or the time spent there is
long enough.  In order to systematically investigate this interference
effect, we have performed double memory experiments in which we have
varied the parameters $T_{2}$ and $t_{w2}$ of the aging stage at the
lower temperature, but kept the initial aging temperature $T_{1}$= 12
K and wait time $t_{w1}$=3 $hrs$ fixed.

\begin{figure}
\vskip 2.0cm
\resizebox{0.48\textwidth}{!}{%
\includegraphics{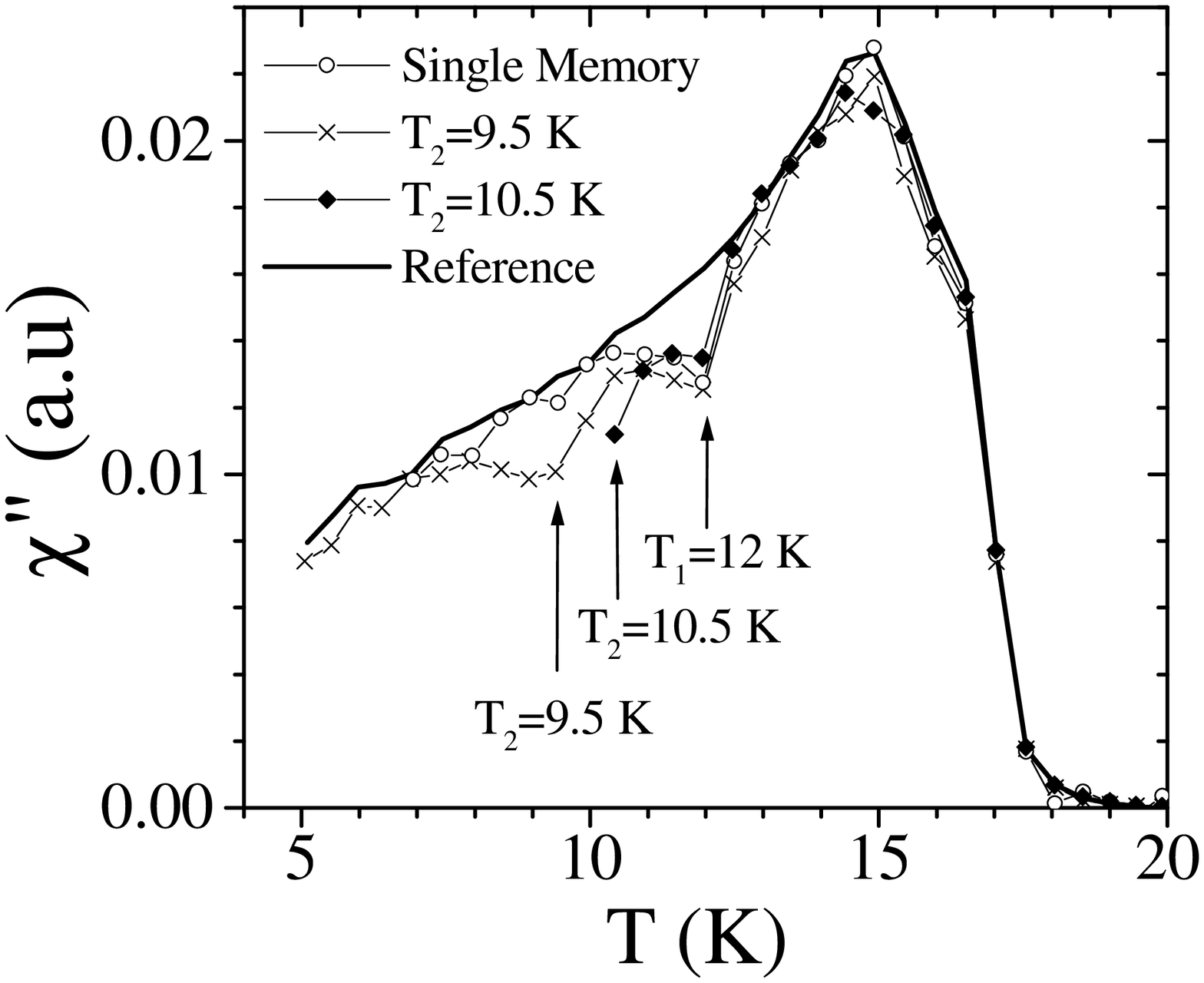} }
\caption{ Effect on the memory at $T_{1}$=12 K ($t_{w1}$=3 $hrs$) of
aging at slightly lower temperatures $T_{2}$ (9.5 K and 10.5 K) during
$t_{w2}$=6 $hrs$. All data shown was taken on reheating after the
various histories. In addition to the reference curve (solid line), a
`single memory' curve (open circles) is presented; it shows the 12K
memory when no additional aging at $T_2$ is performed. For $T_2$=9.5 K
(crosses), there is almost no effect; for $T_2$=10.5 K, the 12K memory
is partly erased.
}
\label{quatre}
\end{figure}

Fig. 4 shows the results using a fixed value of $t_{w2}$=6 $hrs$ but
two different values of $T_2$ (9.5 and 10.5 K).  Two reference heating
curves are added for comparison; one is recorded after a cooling
procedure where no halts are made, and the other after cooling with
only a single halt at $T_1$=12 K for 3 $hrs$ (`single memory').  This
latter curve is a reference for a pure memory effect at 12 K.  The
obtained 12 K dip is about the same for the pure memory and the double
memory curve with $T_{2}$=9.5 K.  However, in the experiment performed
with $T_{2}$=10.5 K , the memory of the dip achieved at $T_{1}$=12 K
has become more shallow.  Thus, for a temperature difference $\Delta
T$=1.5 K, the memory of aging gets partly re-initialised, while for
$\Delta T$=2.5 K no re-initialisation is observed.

\begin{figure}
\vskip 2.0cm
\resizebox{0.48\textwidth}{!}{%
\includegraphics{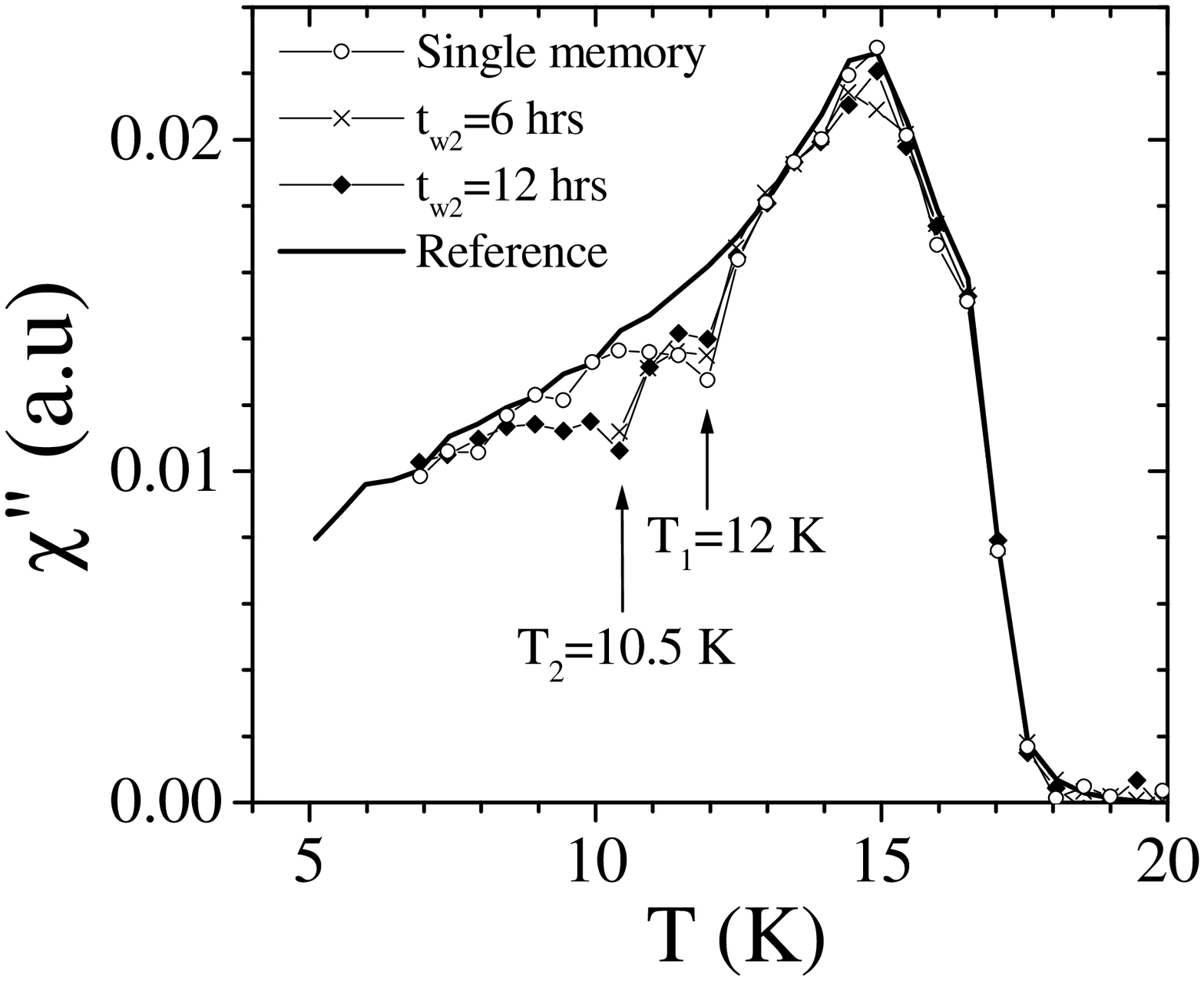} }
\caption{ Effect on the memory at $T_{1}$=12 $K$ ($t_{w1}$=3 $hrs$) of
aging at $T_{2}$=10.5 K during $t_{w2}$=6 (crosses) and 12 $hrs$ (full
diamonds). Same conventions as in Fig.4.
}

\label{cinq}
\end{figure}

Fig.  5 shows the results using a fixed value of $T_{2}$=10.5 K but two
different values of the wait time at $T_2$ $t_{w2}$=6 $hrs$
and $t_{w2}$=12 $hrs$ .  The reference curves are the
same as in Fig.  4.  The longer the time spent at $T_{2}$, the larger
is the part of the memory dip at $T_{1}$ that has been erased.

\section{Discussion}
\label{discu}

\subsection{Memory and chaos effects in phase space pictures}

Some effects related to these memory and memory interference phenomena
have been explored in the past through various experimental
procedures, as well in ac as in dc (magnetisation relaxation following
a field change) measurements \cite{hierarki,Sitges,UppsDT}.  The
hierarchical phase space picture \cite{hierarki} has been developed
as a guideline that accounts for the various results of the
experiments.  Although this hierarchical picture deals with metastable
states as a function of temperature, it is obviously reminiscent of
the hierarchical organization of the pure states as a function of
their overlap in the Parisi solution of the mean field spin glass
\cite{mpv}.  We want to  recall that some more quantitative analyses of the
experiments \cite{sacorbach} have shown that the
barrier growth for decreasing temperatures should be associated with a
divergence of some barriers at any temperature below $T_g$.  Thus,
what can be observed of the organization of the metastable states
might well be applicable to the pure states themselves.  From a
different point of view, another link between the hierarchical picture
and mean field results has been proposed in a tree version of
Bouchaud's trap model \cite{jpbdean}.

The restart of aging when the temperature is again decreased after
having been halted at some value indicates that the
free-energy landscape has been strongly perturbed.  The metastable states
have been reshuffled, but not in any random manner, because the
memory effect implies a return to the previously formed landscape
(with the initial population distribution) when the temperature is
raised back.  The restart of aging then corresponds to the growth of
the barriers and to the birth of other ones, which subdivide the
previous valleys into new ones where the system again starts some 
ab initio aging.  This hierarchical ramification is easily reversed
to produce the memory effect when the temperature is increased back.

The full memory effect seen in the experiment of Fig.  2 requires a
large enough temperature separation $\Delta T=T_1-T_2$, as is shown
from the memory interference effects displayed in Figs.  4 and 5.
If one forgets the restart of aging at the lower temperature, the memory
effect can be given a simple explanation: the slowing down related to
thermal activation is freezing all further evolution of the system.
However, the memory effect takes place while important relaxations
occur at lower temperatures, and it is clear that there must be some
smaller limit of $\Delta T$ below which the $T_2$ evolution is of
influence on the $T_1$ memory.  From other measurements
\cite{hierarki,Sitges}, it has been shown that in the limit of small
enough $\Delta T$'s (of order 0.1-0.5 K), the time spent at $T_2$
contributes essentially additively to the aging at $T_1$, as an
effective supplementary aging time.  In that situation of small
$\Delta T$, the landscape at $T_2$ is not very different from that at
$T_1$ (large `overlap'); the same barriers are relevant to the aging processes,
although being crossed more slowly at $T_2$.  

But this is not the case
in Figs.  4 and 5, where intermediate values of $\Delta T$ have been
chosen.   
The memory interference effect
demonstrated in Figs.  4 and 5 is in agreement with earlier ac and dc
experiments which used negative temperature cycling procedures
\cite{hierarki,UppsDT} and intermediate magnitudes of $\Delta T$.
Such experiments were performed so that the sample first was aged at
$T_1$ a wait time $t_{w1}$, and thereafter cooled to $T_{2}$ and kept
there a substantial wait time $t_{w2}$, after which it was re-heated to
$T_1$, where the relaxation of the ac or dc signal was recorded.  The
results of these experiments are that a partial re-initialisation of the
system has occurred, but that simultaneously a memory of the original aging at $T_{1}$ remains.  In Figs.  4 and
5, the partial loss of the $T_1$ memory dip corresponds to such a
partial reinitialisation.

In this case of an intermediate value of $\Delta T\sim 1$ K, there are
indeed differences between the landscapes at both temperatures.
Still, they are hierarchically related, since a memory effect is
found.  But the memory loss of Figs. 4 and 5 suggests that the 
free-energies of the bottom of the valleys are different at $T_1$ and
$T_2$, meaning that the thermodynamic equilibrium phase is different
from one temperature to another.  As discussed previously
\cite{uppsalaSaclay}, the restart of aging when the temperature is
lowered is suggestive of chaos between the equilibrium correlations at
different temperatures.  The conclusion from the current results is
thus that the free-energies of the metastable states  vary
chaotically with temperature, which  reinforces the idea of a
`chaotic nature of the spin glass phase' \cite{braymoore}.

\subsection{Towards an understanding of memory and chaos effects in
real space}

While these phase space pictures allow a good description of many
aspects of the experimental results, a correct real space picture  would
form the basis for a microscopic understanding of the
physics behind the phenomena.  As aging proceeds, $\chi$'' decreases,
which means a decrease of the number of dynamical processes that have
a time scale of order $1/\omega$.  Aging corresponds to an overall
shift of a maximum in the spectrum of relaxation times towards longer
times, as was understood from the early observations of aging in
magnetisation relaxation experiments \cite{lundgren83}.  Thinking of
the dynamics in terms of groups of spins which are simultaneously
flipped, longer response times are naturally associated with larger
groups of spins.  In such `droplet' \cite{FH} and `domain' \cite{Henk}
pictures, aging corresponds to the progressive increase of a typical
size of spin glass domains.  Difficulties are encountered in this real
space description with `memory and chaos' effects
\cite{uppsalaSaclay}.  On the one hand, the restart of aging
processes when the temperature is lowered indicates the growth of
domains of different types at different temperatures.  On the other
hand, the memory of previous aging at a higher temperature can be
retrieved; thus, the low temperature growth of domains of a given type
does not irreversibly destroy the spin structures that have developed
at a higher temperature.

However, a heuristic interpretation of aging in spin glasses in terms
of droplet excitations and growth of spin glass equilibrium domains,
along the lines suggested in \cite{uppsalaSaclay} and developed in
\cite{jonssonetal}, is perhaps able to include the memory phenomena
discussed in this paper.  The carrying idea of this phenomenology is
that at each temperature there exists an equilibrium spin glass
configuration that is two fold degenerate by spin reversal symmetry.
The simple picture of Fisher-Huse \cite{FH}, where only compact
domains are considered, is hard to reconcile with the memory effect
reported here \cite{uppsalaSaclay}. As suggested in various
theoretical work \cite{fractaldiv,fractclust,jpbdean,mfnoneq}, we
assume that the initial spin configuration results in an
interpenetrating network of fractal `up' and `down' domains of all
sizes separated by rough domain walls. We furthermore propose to
modify somewhat the original interpretation of the `overlap length'
$L_{\Delta T}$.  The standard picture states that the equilibrium
configurations corresponding to two nearby temperatures $T$ and
$T\pm\Delta T$, are completely different as soon as one looks at a
scale larger than $L(\Delta T)$. However, in a non-equilibrium
situation, we believe that some fractal large scale (larger than
$L_{\Delta T}$) `skeletons', carrying robust correlations, can survive
to the change of temperature, and are responsible for the memory
effects. An assumption of this sort is, we think, needed to account
for the existence of domains of all sizes within the initial
condition. If the initial spin configuration was purely random as
compared to the equilibrium one, then the problem would be tantamount
to that of percolation far from the critical point, where only small
domains exist.

The allowed excitations in this system
are droplets of correlated `up' or `down' regions of spins of all
sizes.  Within this model, the magnetisation of the sample in response
to a weak magnetic field is caused by polarisation of droplets, and the
out-of-phase component of the susceptibility directly reflects the
number of droplet excitations in the sample with a relaxation time
equal to $1/\omega$. The size of spin glass domains is in the following
denoted $R$ and the size of a droplet excitation $L$. As a function of time,
the size of the excited droplets grows as $L(T,t_a)$.

The effect of a droplet excitation on the spin configuration is
different depending on the size and position of the droplet and the
age of the spin glass system.  A small droplet excitation $L\ll R$
most probably is just an excitation within an equilibrium spin glass
configuration, yielding no measurable change of the spin system.  An
excitation of size $L\approx R$ may
(i) remove the circumventing domain wall separating an up domain from a
down domain, (ii) slightly displace an existing domain wall or (iii)
just occur within an equilibrium spin configuration.

After a few decades in time, the result of the numerous dispersed
droplet excitations of sizes $L\leq L(T,t_{a})$ is that most domains
of size $R\ll L(T,t_{a})$ are removed, whereas most larger domains remain
essentially unaffected, only having experienced numerous slight domain
wall displacements.  In other words, in this picture, the structure of the
large domains is unaffected by the dynamics. If the temperature now is
changed to a
temperature where the overlap length, $L_{\Delta T}$, is smaller than the
typical droplet size $L(T_1,t_{a})$ active at the original temperature
$T_{1}$, the new
initial condition still leads to an interpenetrating network of up and down
domains of all sizes relevant to the new temperature (chaos implies that
the domain spin structure is different from the equilibrium configuration
at $T_{1}$).  A similar
process as discussed above now creates a spin configuration of
equilibrium structure on small length scales, where only small domains
(up to a size $L(T_2,t_{a})$) are erased, leaving large ones essentially
unaffected. Returning to the original temperature, there will exist a new
domain
pattern on small length scales corresponding to droplets with short
relaxation time and there will additionally remain an essentially
unperturbed domain pattern on large length scales originating from the
initial wait time at this temperature.  The small length scale domains
are rapidly washed out and a domain pattern equivalent
to the original one is rapidly recovered.

These dynamic features of the spin structure are reflected in the
susceptibility experiments discussed above.  The out of phase
component gives a measure of the number of droplets that have a
relaxation time equal to the observation time of the experiment,
$1/\omega$, the decrease of the magnitude of the susceptibility with
time implies that the number of droplets of relaxation time $1/\omega$
decays toward an equilibrium value obtained when `all' domains of this
size are extinguished and all subsequent droplet excitations of this
size occur within spin glass ordered regions.  The fact that a dip
occurs when the temperature is recovered mirrors that the long length
scale spin configuration is maintained during an aging period at
lower temperatures, where only droplet excitations on much smaller
length scales are active.  The memory interference effects are within
this picture immediate consequences of that the droplet excitations at
the nearby temperature $T_{2}$ are allowed to reach the length scales
of the original domain growth at $T_{1}$ leading to an enhanced number
of droplet excitations of the size of this reconstructed domain
pattern.  When heating above the temperature for the aging process,
Fig.  3, and outside the region of overlapping states, the
equilibration processes at the high temperature reach longer length
scales than at the aging temperature, and the memory of the
equilibration at $T_{1}$ is rapidly washed out.  On the other hand, in
the experiments where the sample is cooled (Figs.  4 and 5) below
$T_{1}$, the processes require longer aging time to reach the length
scales of the aging at $T_{1}$ and the intereference effects become
larger with increased time and higher temperature.

\section{Conclusions}
\label{conc}

When cooling a spin glass to a low temperature in the spin glass
phase, a memory of the specific cooling sequence is imprinted in the
spin configuration and this memory can be recalled when the system is
continuously re-heated at a constant heating rate
\cite{uppsalaSaclay}.  E.g. in a `double memory experiment' two
intermittent halts, one at $T_{1}$ a time $t_{w1}$ and another at
$T_{2}$ a time $t_{w2}$ are made while cooling the sample. Depending
on the parameters $T_{2}$ and $t_{w2}$ it is possible to partly
reinitialise (erase) or fully keep the memory of the halt at the higher
temperature $T_{1}$.

These memory and memory interference effects can on the one hand be
incorporated in hierarchical models for the configurational energies
at different temperatures including a chaotic nature of the spin glass
phase.  On the other hand, a preliminary phenomenological
real space picture has been proposed  to account for the observed
phenomena. In our mind, much remains to be done on the theoretical side to
put this `fractal' droplet picture on a firmer footing.

\section{Acknowledgments}

Financial support from the Swedish Natural Science Research Council
(NFR) is acknowledged.
We are grateful to L.F. Cugliandolo, T. Garel, S. Miyashita, M. Ocio and
H. Takayama for valuable discussions, and to L. Le Pape for his
technical support.

\end{document}